\documentclass[11pt,a4paper]{article}
\usepackage{times,latexsym}
\usepackage{url}
\usepackage[T1]{fontenc}
\usepackage{tipa} 

\usepackage[table]{xcolor} 

\usepackage[acceptedWithA]{tacl2018v2}

\usepackage{xspace,mfirstuc,tabulary}

\newif\iftaclinstructions
\taclinstructionsfalse 
\iftaclinstructions
\renewcommand{\confidential}{}
\renewcommand{\anonsubtext}{(No author info supplied here, for consistency with
TACL-submission anonymization requirements)}
\newcommand{\instr}
\fi

\iftaclpubformat 

\else

\fi

\title{Coarse-to-Fine Memory Matching for Joint Retrieval and Classification}

\author{
 Allen Schmaltz \and Andrew Beam \\
 Harvard University \\
  {\sf aschmaltz@hsph.harvard.edu, andrew\_beam@hms.harvard.edu} \\
}

\date{}

\newcommand{\greysmallrule}{\arrayrulecolor{black!30}\cmidrule(l{15em}){1-2}\arrayrulecolor{black}}
\definecolor{QueryColor}{HTML}{941751}
\usepackage{multirow}
\usepackage{basecommon}
\usepackage{graphicx}
\usepackage{booktabs}
\usepackage{array}
\newcolumntype{P}[1]{>{\centering\arraybackslash}p{#1}}
\newcolumntype{T}[1]{>{\raggedright\arraybackslash}p{#1}}

\definecolor{lightestgray}{gray}{0.85}

\begin{document}
\maketitle
\begin{abstract}
  We present a novel end-to-end language model for joint retrieval and classification, unifying the strengths of bi- and cross- encoders into a single language model via a coarse-to-fine memory matching search procedure for learning and inference. Evaluated on the standard blind test set of the FEVER fact verification dataset, classification accuracy is significantly higher than approaches that only rely on the language model parameters as a knowledge base, and approaches some recent multi-model pipeline systems, using only a single BERT base model augmented with memory layers. We further demonstrate how coupled retrieval and classification can be leveraged to identify low confidence instances, and we extend exemplar auditing to this setting for analyzing and constraining the model. As a result, our approach yields a means of updating language model behavior through two distinct mechanisms: The retrieved information can be updated explicitly, and the model behavior can be modified via the exemplar database.   
\end{abstract}

\section{Introduction}

Research over the past two years has transformed the notion of interpreting and updating neural networks in classification settings, via non-parametric (distilled) memories formed by strategically cutting parametric networks \citep{Schmaltz-2019-BLADE-1,Schmaltz2020-multiBLADE}. With exemplar auditing (non-parametric memory matching), we can introspect a model by forming mappings between relevant features and their counterparts in the ground-truth training data, which is a useful mechanism for model and data analysis. Interestingly, in some settings, we can also update model behavior with this mechanism, without re-training, in the sense of updating the labels associated with the test-training exemplar mappings. In this limited sense, neural networks have nudged slightly in the direction of becoming flexible, updatable objects, as opposed to immutable objects dependent on SGD-style updates.

Many practical, real-world NLP tasks can be cast as binary or multi-label classification, and can thus directly make use of the aforementioned analysis and model updating approaches. However, such sequence-oriented classification tasks are but a subset of a larger family of \textit{retrieval-classification} tasks, such as fact verification and question answering, which require an auxiliary step to retrieve information from external datastores, from which the final classification decision is derived. This involves matching over multiple sequences and is typically approached with separate models for the retrieval and classification steps.

The success of exemplar auditing in the basic classification setting, in the sense of effectively matching to relevant features, may be attributed to the distilled memories being formed from the final layers of the network, after the input has been composed. In such settings, once we know where to cut the network, the matching step is comparatively easy, with the resulting analysis use cases not dependent on subsequent deep composition of the matched vectors. However, in the more general case of matching to external datastores in retrieval-classification tasks across multiple sequences, we must confront the \textit{bi-encoder vs. cross-encoder dilemma}: When matching representations from the top layers of a deep network, it is typically more effective to compose all relevant sequences in the input to the network (cross-encoder), but this is computationally infeasible for large datastores, so instead we must resort to matching against representations from independent passes through the network (bi-encoder) \citep[][inter alia]{ReimersAndGurevych-2019-SBERT-sent-sim,UrbanekEtAl-2019-TextAdventure-Bi-vs-Cross-Comparison,HumeauEtAl-2020-Polyencoders}. For large datastores, these challenges are typically addressed with multiple models and re-ranking. We instead propose a single model for retrieval and classification, using a coarse-to-fine search procedure, enabled by adding memory layers to the pre-trained Transformer \citep{VaswaniEtAl-2017-Transformer} language model\footnote{We follow current usage in referring to \textsc{BERT} as a ``language model'', but note that as a result of its imputation-style loss function, it does not in itself yield a well-formed probability over a sequence \cite[c.f.,][]{ChenAndGoodman1999-LMSmoothing}, the traditional notion of a language model in NLP.} \textsc{BERT} \citep{DevlinEtAl-2018-BERT}.

The Fact Extraction and VERification (FEVER) dataset \citep{ThorneEtAl-2018-FEVER-dataset} is a useful test-bed for examining these issues, as it strikes a practical balance between approaching real-world retrieval-classification tasks and having sufficient synthetic simplicity to approach at current computational scales, while also having a large literature of previous works for comparison, including a hidden test set. Large parameter, multi-model pipelines, sometimes also including external linguistic tools, have shown steady gains on the task since the original shared tasks \citep{ThorneEtAl-2018-FEVER-SharedTask1,ThorneEtAl-2019-FEVER-SharedTask2}. Orthogonal to pushing the ensemble SOTA, works such as \citet{LeeEtAl-2020-LM-as-KB} have used FEVER to analyze the extent to which the parameters of LMs, themselves, can serve as a knowledge base. Additionally, recent work has suggested that typical evaluations on synthetic tasks, such as FEVER, may be misleading, or at least not express the full effectiveness picture, due to idiosyncrasies in data construction, allowing models to exploit spurious features \citep{Schuster-Etal-2019-FEVER-symmetric-bias}.

In summary, we make the following contributions in this work:

\begin{itemize}
\item We present an approach for learning and inference with a single language model for both retrieval and classification. This allows us to leverage the learned knowledge captured by the LM's parameters, as well as explicit knowledge provided in retrieval. Accuracy on FEVER fact verification exceeds all of the original 2018 Shared Task systems; significantly exceeds approaches that only rely on the Transformer LM's parameters as a knowledge base; and approaches some multi-pipeline systems despite using significantly fewer parameters and no external linguistic tools.
\item To achieve this, we introduce a novel coarse-to-fine search procedure across non-parametric memories dynamically created by using the LM in both a bi-encoder and cross-encoder manner. This is used for inference and for training, making it possible to backpropagate through both retrieval and classification using a supervised similarity loss.
\item Additionally, we demonstrate how the non-parametric distances from tightly coupled retrieval and classification can be leveraged to identify low-confidence instances. We also extend exemplar auditing to this setting, adding an additional means for analyzing the model and constraining the output. 
\end{itemize}

\section{Task}

The FEVER task aims to assess whether a claim (a short, declarative sentence) is true, false, or unverifiable, based on one or more sentences from Wikipedia. At inference, a model is given a claim and must retrieve applicable sentences from Wikipedia, with results evaluated on the 3-class classification accuracy, as well as a ``FEVER score'', which only counts as correct those predictions whose retrieved sentences (up to five) include at least one complete covering of ground-truth associated evidence sentences. The Wikipedia sentences are uniquely described by the Wikipedia article title and the sentence number. Appendix~\ref{sec:AppendixSearchExample} provides an example of a claim and associated evidence.

The final test set has hidden labels, and is evaluated on CodaLab.\footnote{Fact Extraction and VERification (FEVER) Challenge: \url{https://competitions.codalab.org/competitions/18814}. Our results were submitted to CodaLab on October 18, 2020.} 

\paragraph{Symmetric Dataset} We also consider the 2-class (true, false) analysis sets of \citet{Schuster-Etal-2019-FEVER-symmetric-bias}, which only considers claims with single sentence evidence, to analyze hypothesis bias. This set re-annotates some of the claims to flip the decision label, and also re-annotates the corresponding evidence sentences to match that re-labeling, and vice-versa (hence, ``symmetric''). As a result, this set also serves as a proxy for the setting of updating the datastore (here, changing the underlying Wikipedia sentence).

\section{Methods}

The iterative, coarse-to-fine search procedure is enabled by a model consisting of a single Transformer combined with three independent single-width-kernel, multiple-filter CNNs that serve to create ``memory'' vectors. Each of the CNNs takes as input the shared top layer of the Transformer (for each input WordPiece) concatenated with word embeddings, the latter of which are unique for each CNN. A max-pool operation over each CNN creates a memory vector, which is used either as a query, or as part of a database, depending on the input sequence. These memory vectors are used to perform matching between sequences, in training and at inference.

This mechanism is used in both a bi- and cross- encoder manner. To aid the presentation, we refer to each of the three CNNs as a ``level'', as each serves a distinct role in the coarse-to-fine search procedure, with the candidate input sequences of the latter levels dependent on the results of the prior levels.\footnote{As a point of contrast, we do not use the term ``layer'' in this context, as each of the CNNs similarly serves as the final layer after the base Transformer.} Each level takes as input a \textsc{query} sequence and a \textsc{support} sequence, with the top-$k$ matched memory vectors determining the candidate \textsc{support} sequences for the subsequent levels. For FEVER, a \textsc{query} sequence is a claim, and a \textsc{support} sequence contains one or more relevant Wikipedia sentences, and for some levels, additionally a classification label and a copy of the claim. 

Specifically, each token $t^q_1,\ldots,t^q_n,\ldots,t^q_N$ in the \textsc{query} sequence for each level is represented by a $D$-dimensional vector, where $N$ is the length of the sequence, including padding symbols, as necessary. $D$ is the concatenation of randomly initialized word embeddings and the top output Transformer layer corresponding to that token. The \textsc{support} sequence is analogous: $t^s_1,\ldots,t^s_n,\ldots,t^s_N$.

The convolutional layers are then applied to the $\reals^{D \times N}$ matrix for each sequence, using a filter of width 1, sliding across the unigrams of the input. The convolution results in a feature map $\boldh_m \in \reals^{N}$ for each of $M$ total filters. 

We then compute $g_m = \max (\boldh_m)$,
%
%
a max-pool over the unigram dimension resulting in $\boldg \in \reals^M$. This is calculated for both the \textsc{query} sequence, $\boldg^q$, and the \textsc{support} sequence, $\boldg^s$. These dense representations of the input sequences are then used for search and learning, as described further below.

\subsection{Coarse-to-Fine Search}

We seek to associate a \textsc{query} sequence with its corresponding \textsc{support} sequence. For each $\boldg^q$, we search for the $\boldg^s$ that minimizes $\lVert \boldg^q-\boldg^s \rVert_2$, the Euclidean distance between the memory vectors of the \textsc{query} and \textsc{support} sequences.

\begin{figure*}[hbt!]
    \centering
    \includegraphics[scale=0.50]{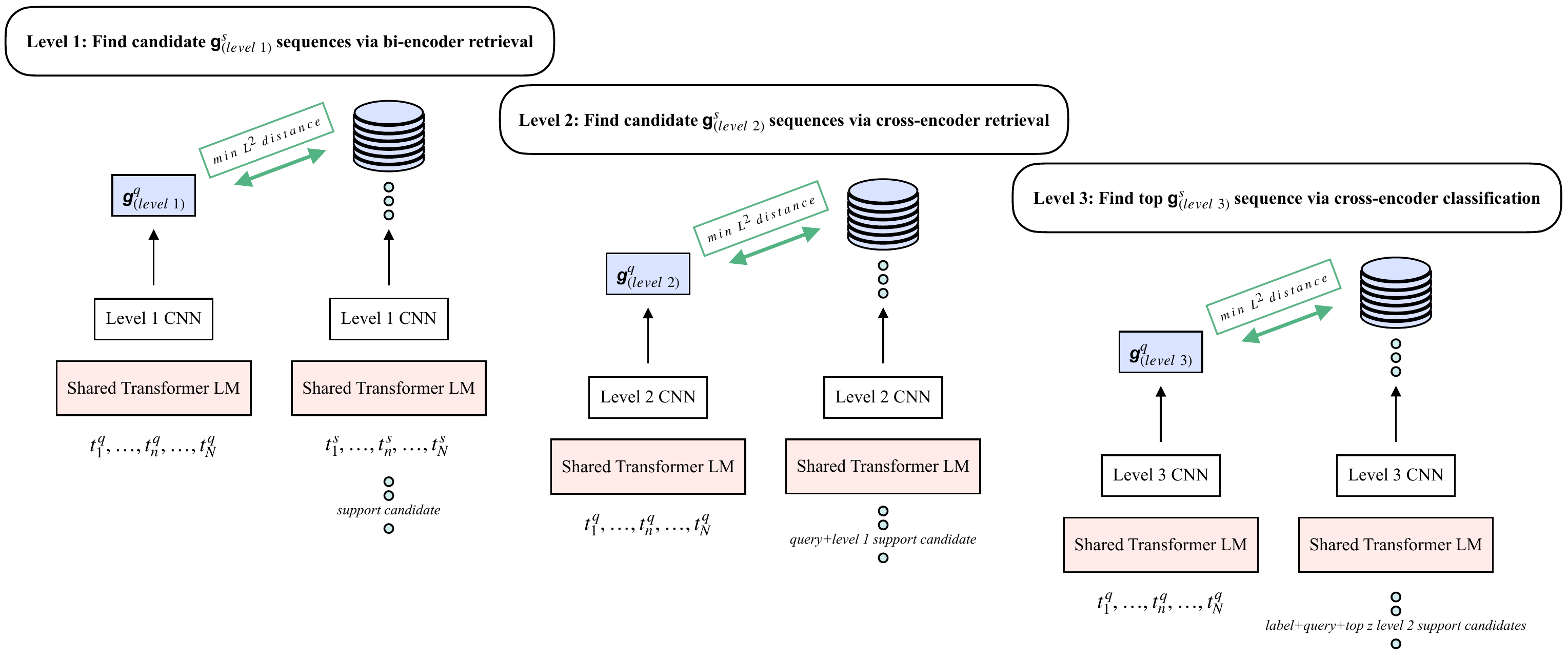}
    \caption{Coarse-to-fine joint retrieval and classification search with a shared language model and independent memory layers. Candidate \textsc{support} sequences are progressively winnowed with bi-encoder (level 1) and then cross-encoder (level 2) retrieval. Final classification of a \textsc{query} sequence is determined by a marginalization over the top-$z$ candidate sequences from the previous level via cross-encoder (level 3) classification. Parameter drift of the shared Transformer is avoided by jointly backpropagating gradients through all three levels during training.}
    \label{fig:Coarse-to-fine-search}
\end{figure*}

The search procedure starts with a coarse, but relatively efficient, bi-encoder matching over all sequences, followed by a more expensive, but also more accurate, cross-encoder matching over a subset of the winnowed sequences from the previous level. A final level handles classification by incorporating the classification labels as part of the \textsc{support} sequences. Figure \ref{fig:Coarse-to-fine-search} illustrates this search process.

\paragraph{Level 1 (bi-encoder retrieval)} The first level performs a coarse bi-encoder search to match \textsc{query} sequences with the nearest \textsc{support} sequences. A \textsc{query} sequence is a FEVER claim, and a \textsc{support} sequence in level 1 is a single Wikipedia sentence. Only the top-$k_1$ \textsc{support} sequences are considered in level 2.   
 
\paragraph{Level 2 (cross-encoder retrieval)} Level 2 performs a finer grained search, only considering the subset of \textsc{support} sequences filtered in the first level. This is cross-encoder matching, with the \textsc{support} sequences constructed from a concatenation of the FEVER claim with each candidate Wikipedia sentence that survived level 1. In this way, the \textsc{support} sequences are unique for each \textsc{query} sequence, so this is only computationally feasible with the candidate winnowing of level 1. Only the top-$z$ \textsc{support} sequences are then considered in level 3.

\paragraph{Level 3 (cross-encoder classification)} The final level performs classification, marginalizing over the top-$z$ \textsc{support} sequences from level 2 in the following sense. The \textsc{support} sequences of level 3 consist of the claim concatenated with the top-$z$ Wikipedia sentences from level 2. Unique \textsc{support} sequences are then created by concatenating the classification label (true, false, or unverifiable). The matching of the \textsc{query} sequence to the \textsc{support} sequence then determines the final classification label. In other words, the search in level 3 is for the nearest label, with the evidence sentences determined by level 2.

Separately, independent of search, for training only, we also consider \textsc{support} sequences in level 3 constructed with the reference Wikipedia sentences. When the distinction is not otherwise clear, we refer to these sequences as level 3 \textit{reference} sequences, and the level 3 sequences constructed from search as level 3 \textit{prediction} sequences.

\paragraph{Sequence Input Formatting} With the intuition of helping the model distinguish between the various levels (retrieval, coarse or fine, and classification, predicted and reference), we also add distinct prefixes to the \textsc{query} sequences in levels 2 and 3 (separately for reference and prediction sequences in level 3), and the \textsc{support} sequences in level 2. In all cases, we prepend the article title and sentence index to each Wikipedia evidence sentence. Table~\ref{tab:search-example} in Appendix~\ref{sec:AppendixSearchExample} illustrates the encoding of the sequences across levels for an example from the Dev set. 

\subsection{Training} Unlike prior work with pipeline systems with separate models for retrieval, sentence selection, and classification, our unified model is trained end-to-end. Across all three levels, we seek to minimize the distance between the memory vectors of \textsc{query} and \textsc{support} sequences of correct matches and maximize the distance between the memory vectors of incorrect matches. 

Specifically, we calculate $\bdelta=\left|\boldg^q-\boldg^s\right| \in \reals^M$, the absolute value of the difference between the max-pool filter values across the \textsc{query} and \textsc{support} sequences (the ``difference vector''). We then seek to minimize the BCE loss
\begin{align*}
L_m &= -Y\cdot \log \sigma(\delta_m) - (1-Y)\cdot \log (1-\sigma(\delta_m)), 
\end{align*}
where $Y=0$ if $\bdelta$ is associated with a correct \textsc{query} and \textsc{support} match, and $Y=1$ otherwise. The loss is averaged across all filters $M$, over all 3 levels, over all \textsc{query} and \textsc{support} pairs in a given mini-batch. In effect, we are able to backprop jointly across retrieval and classification. 

We informally refer to our proposed loss as a supervised similarity loss, as opposed to a contrastive loss, as used in vision \citep[e.g.,][]{ChenEtAl-2020-SimCLR}, since there is no notion of data augmentation via local data perturbations. Instead, our loss is enabled by our novel coarse-to-fine search mechanism, with positive and negative instances constructed as follows.

\paragraph{Positive Examples} For levels 1 and 2, correct matches are constructed from the ground-truth retrieval data. For level 3, we consider the ground-truth Wikipedia sentences when constructing reference \textsc{support} sequences for verifiable claims. Additionally, when constructing level 3 prediction sequences, we combine the correct label (true, false, unverifiable) with the Wikipedia sentences from the top-$z$ \textsc{support} sequences from level 2. This latter case emulates inference, as the correct Wikipedia sentences will not necessarily be in these \textsc{support} sequences at training, even if the label is correct.\footnote{While typical multi-model pipeline systems are trained by bootstrapping the unverifiable instances by generating evidence sentences from another model, the coarse-to-fine mechanism naturally handles this case in level 3 with the single model. This points to the potential for adding semi-supervised instances to other levels.}

\paragraph{Negative Examples} For levels 1 and 2, hard negatives are constructed by performing the coarse-to-fine search and retrieving the nearest incorrect matches. For level 3, negatives are constructed by flipping the concatenated label string in the \textsc{support} sequences. 

\subsection{Inference} At test, we perform the coarse-to-fine search, taking as our prediction the label and Wikipedia sentences from the nearest \textsc{support} sequence match in level 3.  

\subsection{Level Distances} From this coupled system, for every traversal through the coarse-to-fine search, we generate relative distances to the selected retrieved information (level 2) and the joint retrieval and classification decision (level 3). We can use this information to analyze and constrain our predictions.

\subsection{Exemplar Auditing via Difference Vectors} 

Previous works have demonstrated effective approaches for creating exemplar vectors for sequence-based classification tasks when the unit of analysis is oriented toward the token level. The situation is more complicated with retrieval-classification tasks, since the final decision is determined by multiple sequences, and the connections between individual tokens and the final decisions are often difficult to consistently articulate for humans. However, the joint nature of our model enables a means of capturing the abstract, compositional patterns that determine the final label decision. From our coupled system, we create exemplar vectors by concatenating the difference vectors, $\bdelta$, from levels 2 and 3. We create a database of these exemplar vectors that we then use to analyze the model by matching to analogous vectors created at inference. Note that this database is distinct from the memories constructed as part of the coarse-to-fine search.

\section{Experiments}\label{sec:experiments}

We examine the behavior of our proposed model on the standard FEVER dataset, submitting to CodaLab for final evaluation on the hidden test set. We further analyze the model in the context of the challenging re-annotation sets of \citet{Schuster-Etal-2019-FEVER-symmetric-bias}. Our code will be made publicly available for replication purposes.\footnote{We use the PyTorch (\url{https://pytorch.org/}) reimplementation of the original BERT code base available at \url{https://github.com/huggingface/pytorch-pretrained-BERT}. Our code will be available no later than publication at \url{https://github.com/allenschmaltz/memory_matching}.}

\subsection{Data}

We use the publicly available FEVER 3-class data for training the model. For training, we only consider one evidence set per claim, preferring smaller evidence sets with sentences near the beginning of the article, and we truncate evidence sets to a max size of 2. Around 88\% of the training set claims have an evidence set consisting of a single sentence. 

We also perform a very coarse, lightweight (model free) filtering of the space of Wikipedia articles under consideration for each claim. After removing parenthetical information and punctuation, and lowercasing, for each claim, we only consider the titles that have the longest lexical cover starting from each token in the claim, left-to-right. In training and dev, this results on average of around 50 articles, or 300 sentences, under consideration for each claim. Around 97\% of the claims have at least 1 ground-truth evidence sentence in their resulting set. 

This preprocessing is sufficient to reduce the retrieval space to be manageable on a single GPU, if level 1 is bi-encoded, but cross-encoding would still be prohibitive. Because there is some sharing across the covered sets, which can be leveraged by bi-encoding, this results in around 625k sentences processed in level 1 for the 145k claims of the training set; 296k sentences in level 1 for the 20k claims of the Dev set; and 329k sentences in level 1 for the 20k claims of the hidden test set. We otherwise do not make any distinctions between the articles, nor do we employ a separate IR system, as used in some previous works. For example, we do not distinguish between regular and disambiguation articles, and instead allow the model to learn the distinctions. 

\paragraph{Symmetric Dataset}

There are two versions of the symmetric re-annotation sets of \citet{Schuster-Etal-2019-FEVER-symmetric-bias} dataset: The version from the published paper, \textsc{sym\textsubscript{gen.}}, and a subsequent version with a dev-test split, \textsc{sym\textsubscript{dev.v2}} and \textsc{sym\textsubscript{test.v2}}, available in the public repo\footnote{\url{https://github.com/TalSchuster/FeverSymmetric}}. We consider the first version for comparison to previous work, and we also consider the new set as it provides a held-out set with which to examine updating the exemplar database. In these 2-class sets, single sentence retrieval is given, and for a subset of instances, the evidence and/or claims have been strategically modified to aid examination of a model's reliance on class conditional distributional characteristics. We use these sets to analyze the model's ability to identify---and predict over---out-of-domain samples. We use the labels \textsc{train\textsubscript{1-evidence}} and \textsc{dev\textsubscript{1-evidence}} to indicate the original FEVER train and dev sets limited to 2-class claims, with given single sentence evidence sets. Because these subsets from previous works have dropped the Wikipedia title and sentence index, we perform a preprocessing step to heuristically re-associate this metadata to the \textsc{support} sequences based on the original FEVER data.

\subsection{Models}\label{sec:Experiments:Models}

Our model, \textsc{BERT\textsubscript{base}+MemMatch}, uses a separate CNN to construct the memory representations for each of the three levels. Each CNN consists of 1000 filters and has a kernel width of 1. Input to each CNN consists of the 768-dimension hidden state of the top layer of \textsc{BERT\textsubscript{base}}, which corresponds to each input WordPiece, concatenated to a randomly initialized 300-dimension word embedding with a vocabulary size of 7,500. We set the max length of level 1 to 50 tokens, level 2 to 100 tokens, and level 3 to 150 tokens.\footnote{Transformers are quadratic in the sequence length, so keeping lengths short instead of using the full 512 max length of \textsc{BERT\textsubscript{base}}, at the cost of a limited amount of truncation of the Wikipedia sentences, is helpful in order to batch sequences from all levels together in training on a single GPU.} 

\paragraph{Training}

Our model differs significantly from previous approaches in that the coarse-to-fine search mechanism allows us to train retrieval and classification together, allowing the sharing of parameters across a single Transformer. At the start of each training epoch, we run the search procedure to find hard negatives and to determine the evidence sentences over which to marginalize for the prediction sequences in level 3. We then backpropagate through all levels by batching all sequences, from all levels, together for each claim when running a backward pass.

We train for up to 15 epochs with $k_1=10$, $z=3$, choosing the epoch with the highest accuracy on the held-out dev set. We then continue training with a larger beam, $k_1=30$, again choosing the epoch with highest held-out accuracy, and terminate training once accuracy has not increased for 4 epochs. Each epoch we alternate freezing either the Transformer parameters or all CNN parameters, which we found works well for learning memory vectors with strong matching effectiveness. In practice, training takes 4-5 days on a single V100 GPU (with 32 GB of memory). Additional training details are in Appendix~\ref{sec:AppendixAdditionalTrainingDetails}.

\paragraph{Inference}

At test, we set $k_1=100$. During training, we keep $z=3$, since the claims have at most 2 reference Wikipedia sentences, allowing us to keep the total sequence lengths relatively short in level 3. At inference, we allow $z$ to increase to 5 for comparison to previous works in terms of the FEVER score, which always considers retrieval sets of up to size 5, regardless of order. 

\paragraph{Exemplar Auditing}

In the analysis of the symmetric dataset, we create the exemplar database from \textsc{train\textsubscript{1-evidence}}, removing unverifiable claims. When running inference over \textsc{sym\textsubscript{test.v2}}, we additionally add \textsc{sym\textsubscript{dev.v2}} to the exemplar database.
 
\subsection{Existing Models for Comparison}\label{sec:ExistingModels}

Existing models for FEVER are highly heterogenous. Reliable empirical comparisons are aided by the hidden test set, but meaningful comparisons are greatly complicated by differing parameter counts, efficiencies, and learning signals.

Typical approaches have separate models for document retrieval, sentence selection, and label classification. The pipeline system of \citet{NieEtAl-2019-NSMN-FEVER-2018-SOTA}, \textsc{NSMN}, using semantic matching networks, was the highest ranking system in the original FEVER Shared Task in 2018. Subsequent work \citep{NieEtAl-2020-NSMN-style-label-sharing} examines the impact of parameter and data sizes, proposing a compounded label approach to unify sentence selection and classification, but using the separate method of \citet{NieEtAl-2019-NSMN-FEVER-2018-SOTA} for document selection. We use the label \textsc{CompoundLabel} for this work; it provides a reference point for a smaller parameter system, albeit not an end-to-end system.

Recent work has further pushed the SOTA, combining large multi-model Transformer models with graph networks, \textsc{GEAR} \citep{ZhouEtal-2019-GEAR}, and graph-based reasoning over semantic role labeling, \textsc{DREAM} \citep{ZhongEtal-2020-DREAM}, which is the current SOTA.

\citet{LeeEtAl-2020-LM-as-KB} carefully examines the ability of language models to serve as knowledge bases, without supervised evidence training, providing baselines with a standard frozen \textsc{BERT\textsubscript{large}} model, fine-tuning the Transformer parameters (\textsc{BERT\textsubscript{large}+FT}), and a new model that freezes \textsc{BERT\textsubscript{large}} and extracts features with strategic named-entity masking for input into an entailment model (\textsc{BERT\textsubscript{large}+KBfeat}).

For reference, we also consider the contemporaneous \textsc{RAG} model \citep{LewisEtAl-2020-RAG}, a generative sequence model with a non-parametric retrieval component. It is not directly comparable to our work and is conceptually and architecturally distinct: \textsc{RAG} jointly fine-tunes a pre-trained bi-encoder (with two BERT\textsubscript{base} models, one of which remains frozen during \textsc{RAG} training) for retrieval with a separate, large pre-trained seq2seq model for classification, whereas \textsc{BERT\textsubscript{base}+MemMatch} jointly fine-tunes a single pre-trained \textsc{BERT\textsubscript{base}} model as both the retriever and classifier. \textsc{BERT\textsubscript{base}+MemMatch} has the advantage of supervised FEVER evidence training, but \textsc{RAG} has the advantage of significantly greater parameters, additional amounts of retrieved text for classification, and signal from QA datasets\footnote{\textsc{RAG} is fine-tuned with supervised FEVER claims but not FEVER evidence; however, importantly, the bi-encoder is not pre-trained in an unsupervised manner in the sense of a Cloze-style task. Rather, it is pre-trained with direct and indirect supervision from labeled QA datasets, which also derive \textsc{support} sequences from Wikipedia \citep{KarpukhinEtAl-2020-DPR}.}. It is nonetheless of interest, since except for the claim-only LM models, it is the closest extant comparison to an end-to-end Transformer-based model for FEVER.


Finally, for the symmetric analysis experiments, we consider the model of \citet{Schuster-Etal-2019-FEVER-symmetric-bias}, which fine-tunes a \textsc{BERT\textsubscript{base}} model with a training set re-weighting scheme designed to reduce hypothesis bias. This model (\textsc{BERT\textsubscript{base}+RW}) is a cross-encoder given the ground-truth evidence during training and inference. We also include the \textsc{BERT\textsubscript{base}} baseline from this work. 

\section{Results}

\paragraph{FEVER}

\begin{table}
\centering
\tiny
\begin{tabular}{lccccc}
\toprule
Model & Acc. & FEV. & Pt.ct. & IR & Ling. \\
\midrule
\textsc{GEAR} & 71.60 & 67.10 & $>204$ & $\bullet$ & $\bullet$ \\
\textsc{DREAM} & 76.85 & 70.60 & $\ge\left(373,833\right]$ & $\bullet$ & $\bullet$\\
\midrule
\textsc{CompoundLabel} & 66.21 & 61.65 & 18 & $\bullet$ & \\
\textsc{NSMN} & 68.16 & 64.23 & 28 & $\bullet$ & $\bullet$ \\
\midrule
\rowcolor{lightestgray} \textsc{BERT\textsubscript{large}} & 38. & N/A & 340 & & \\
\rowcolor{lightestgray} \textsc{BERT\textsubscript{large}+FT} & 57. & N/A & 340 & & \\
\textsc{BERT\textsubscript{large}+KBfeat} & 49. & N/A & $>340$ & & $\bullet$ \\
\rowcolor{lightestgray} \textsc{RAG} & 72.5 & N/A & 626 & & \\
\midrule
\rowcolor{lightestgray} \textsc{BERT\textsubscript{base}+MemMatch} & 70.42  & 63.95 & 120 & & \\
\bottomrule
\end{tabular}
\caption[Main results]{FEVER hidden test results, evaluated on CodaLab, in terms of accuracy (Acc.) and FEVER score (FEV.). We include parameter estimates, in millions (Pt.ct.), with ranges when Transformer sizes or auxiliary model details are unspecified in the original publications (see Appendix~\ref{sec:AppendixParameterEstimates} for details). We indicate whether a pipeline model also uses non-neural IR features (IR) and whether external linguistic tools are used (Ling.). We distinguish models that do not train with sentence-level evidence with N/A as the FEVER score. All but the last line are test results reported in previous works.~{\color{black}\colorbox{lightestgray}{Light-gray rows}}~contain end-to-end models.} 
\label{table:test-results-codalab}
\end{table} 

Table~\ref{table:test-results-codalab} shows results across models on the hidden test set, evaluated on CodaLab. Our model is significantly stronger than the end-to-end language models that lack retrieval, and it approaches the accuracy of one of the recent strong multi-model systems, despite having fewer parameters, not using external linguistic tools, and consisting of a single end-to-end model. Additionally, our tightly coupled end-to-end system achieves $97\%$ of the accuracy of \textsc{RAG} using only $19\%$ of the parameters.

The \textsc{BERT\textsubscript{large}} models from \citet{LeeEtAl-2020-LM-as-KB} are important in that they provide rough empirical  ranges on what we might expect from a claim-only LM at the given parameter size. In particular, \textsc{BERT\textsubscript{large}+FT} is much stronger than a random baseline, which would be around 33.00, since the dataset is balanced. Interestingly, by adding around 10 million parameters to the smaller \textsc{BERT\textsubscript{base}} model with our proposed approach with supervised evidence training, we can dramatically improve effectiveness. It is conceivable that LM-only effectiveness will increase with increasing parameters; however, at the scale examined here, the difference between explicitly retrieving Wikipedia sentences  and only relying on LM parameters, which indirectly learn Wikipedia from pre-training, is stark. 

The bar is high for the carefully tuned, bespoke multi-model pipelines, especially the current SOTA, \textsc{DREAM}, which incorporates some of the current best-of-class models for each of document retrieval, sentence selection, and classification. Our model compares favorably to the non-Transformer models, \textsc{CompoundLabel} and \textsc{NSMN}. Our model has additional parameters relative to those baselines; however, our single model potentially lends itself to more general applications, as noted in Section~\ref{sec:discussion}. Interestingly, the accuracy of our model is only about 1 point less than \textsc{GEAR}. We see a larger decrease in the retrieval-oriented FEVER score, which is not surprising, since \textsc{GEAR} has a sophisticated Wikipedia-oriented retrieval system using the MediaWiki API and additional models. 

\begin{table}
\centering
\footnotesize
\begin{tabular}{ccccc}
\toprule
 & \multicolumn{2}{c}{$z=3$} & \multicolumn{2}{c}{$z=5$} \\
$k_1$ & Acc. & FEV. & Acc. & FEV.\\
\midrule
3 & 67.67 & 50.71 & \multicolumn{2}{c}{\textit{(equivalent to $z=3$})}\\
5 & 69.33 & 54.73 & 69.39 & 55.17 \\
30 & 73.04 & 65.27 & 73.07 & 65.93 \\
100 & 73.49 & 66.85 & 73.59 & 67.70 \\
\bottomrule
\end{tabular}
\caption[Main results]{Dev. set accuracy (Acc.) and FEVER score (FEV.) vs. level 1 beam size ($k_1$) and the number of evidence sentences marginalized in level 3 ($z$) for \textsc{BERT\textsubscript{base}+MemMatch}.} 
\label{table:dev-beam-size}
\end{table} 
 
In FEVER, retrieval is important and search is non-trivial. On the one hand, claim-only \textsc{BERT} fine-tuned models do much better than random, suggesting some combination of signal encoded in the parameters and/or hypothesis bias; on the other hand, the distance to \textsc{BERT} models given retrieval evidence is very large ($>10$ points). Furthermore, as seen in Table~\ref{table:dev-beam-size}, level 3 classification in our model benefits from better retrieval from levels 1 and 2, which is controllable via increasing the beam size at the expense of additional compute. For submission to CodaLab, we increase $z$ to 5 at inference, which increases the FEVER score by nearly 1 point, since the FEVER metric penalizes low recall evidence predictions. 

The addition of level 2 (i.e., the cross-encoder for retrieval), rather than only relying on level 1, is important for evidence selection. With the search configuration of our submitted model, a strict, complete evidence match in the top 1 (or up to 2, if the ground-truth has 2 evidence sentences) beam positions rises from 37.8\% of the dev claims in level 1 to 74.1\% in level 2. Similarly, the percentage of \textsc{support} sequences at the top of the beam associated with a correct \textit{document} (i.e., Wikipedia title) rises from 72.2\% in level 1 to 91.5\% in level 2. 

\paragraph{Symmetric Dataset}

\begin{table}
\centering
\tiny
\begin{tabular}{lcccc}
\toprule
 & \multicolumn{2}{c}{\textsc{dev\textsubscript{1-evidence}}} & \multicolumn{2}{c}{\textsc{sym\textsubscript{gen.}}}  \\
Model & Acc. & $n$ & Acc. & $n$ \\
\midrule
\textsc{BERT\textsubscript{base}} & 86.2 & 16,664 & 58.3 & 717 \\
\textsc{BERT\textsubscript{base}+RW} & 84.6 & 16,664 & 61.6 & 717 \\
\midrule
\textsc{BERT\textsubscript{base}+MemMatch} & 92.6 & 16,664 & 60.8 & 717 \\
\textsc{BERT\textsubscript{base}+MemMatch+dist} & 92.1 & 6,488 & 75.8 & 343 \\
\textsc{BERT\textsubscript{base}+MemMatch+dist+exchange} & 92.6 & 16,664 & 69.5 & 717 \\
\textsc{BERT\textsubscript{base}+MemMatch+ExA\textsubscript{tp}} & 93.4 & 14,722 & 61.9 & 641 \\
\bottomrule
\end{tabular}
\caption[Main results]{Results on the analysis subsets of \citet{Schuster-Etal-2019-FEVER-symmetric-bias}. The results in the first two rows are from previous works. Some results only predict a subset of size $n$ of the total claims.}
\label{table:symmetric-results}
\end{table}

\begin{figure}[t]
  \begin{center}
    \includegraphics[width=\columnwidth]{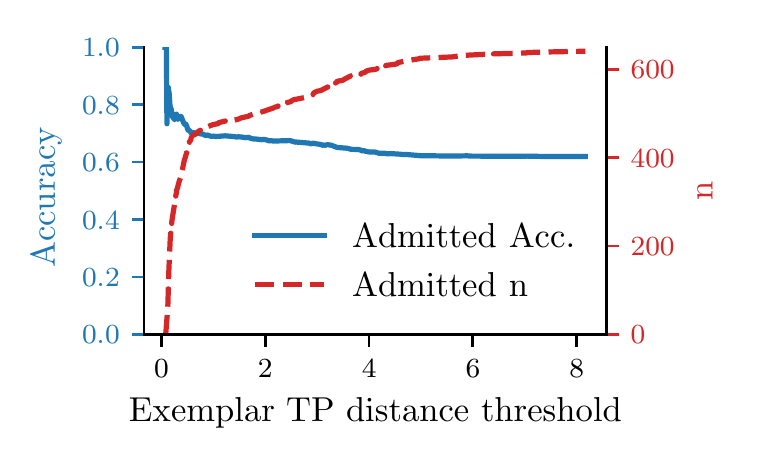} 
    \caption{\label{fig:exa-tp-dist} Accuracy of classification predictions on the out-of-domain \textsc{sym\textsubscript{gen.}} data for \textsc{BERT\textsubscript{base}+MemMatch+ExA\textsubscript{tp}}, when setting a max allowed distance cutoff to the exemplar at which to admit predictions. Instances matched to true positive exemplars with closer distances tend to be associated with more accurate model predictions on this challenging set.}
  \end{center}
\end{figure}

\begin{table}
\centering
\tiny
\begin{tabular}{lcc}
\toprule
 & \multicolumn{2}{c}{\textsc{sym\textsubscript{test.v2}}} \\
Model & Acc. & $n$ \\
\midrule
\textsc{BERT\textsubscript{base}+MemMatch} & 70.8 & 712 \\
\textsc{BERT\textsubscript{base}+MemMatch+dist} & 81.3 & 359 \\
\textsc{BERT\textsubscript{base}+MemMatch+dist+exchange} & 76.4 & 712 \\
\textsc{BERT\textsubscript{base}+MemMatch+ExA\textsubscript{tp}} & 72.8 & 629 \\
\textsc{BERT\textsubscript{base}+MemMatch+ExA\textsubscript{update}} & 78.1 & 712 \\
\bottomrule
\end{tabular}
\caption[Main results]{Results on the \textsc{sym\textsubscript{test.v2}} out-of-domain split.}
\label{table:symmetric-results-testv2}
\end{table} 

The effectiveness of the model on the analysis subsets of \citet{Schuster-Etal-2019-FEVER-symmetric-bias} appears in Table~\ref{table:symmetric-results}. On \textsc{sym\textsubscript{gen.}}, we see that the \textsc{BERT\textsubscript{base}+MemMatch} accuracy is higher than a standard \textsc{BERT\textsubscript{base}} model and only about 1 point less than the training-set re-weighting of previous work, \textsc{BERT\textsubscript{base}+RW}, despite adding only around 10\% (10 million) more parameters and not modifying the training set. However, as with earlier models, the accuracy on this challenging set is dramatically lower than the results on \textsc{dev\textsubscript{1-evidence}}, which has a similar distribution to that seen in training. Breaking this down further, we see that the accuracy on the symmetric set (\textsc{sym\textsubscript{gen.}}) with claims modified to flip the label, with the datastore unchanged, is actually reasonably high, at 81.6 percent. However, the accuracy on claims with modified evidence sentences is very low, at 50.4 percent, which is essentially \textit{no better than chance}, if not making use of the ability of the model to identify outliers. 

Characteristics of the datastore changes (modified evidence sentences) can help explain the low effectiveness. The datastore changes in \textsc{sym\textsubscript{gen.}} can change the type of an entity, as opposed to only changing aspects that would typically change over time, such as occupations or new achievements. For example, a TV show might become a music album. These changes are challenging if the model has already encoded such type information into its parameters. Additionally, modifying Wikipedia sentences in isolation may introduce some subtle artifacts that push the language distribution beyond what is typical of the domain. Table~\ref{tab:retrieval-dist-examples-sym} in Appendix~\ref{sec:AppendixSymmetricAnalysis} provides examples. As a result, this dataset is a useful, challenging dataset with which to examine identifying low confidence outliers, and updating the more abstract model behavior, beyond just modifying the datastore. 

Our coupled system produces distances between sequences at each level. In Table~\ref{table:symmetric-results}, we include the accuracy (\textsc{BERT\textsubscript{base}+MemMatch+dist}) when only admitting predictions for which the level 2 and level 3 distances are less than the mean of the distances for correct retrieval and correct label predictions, respectively, seen in training. At the cost of emitting null predictions, this results in essentially no change in accuracy on the dev set, while dramatically increasing the accuracy by 10 points on \textsc{sym\textsubscript{gen.}}. In this case, the accuracy on the instances with changed evidence sentences rises to 66.2 (out of 204 admitted predictions, which is 43\% of all such claims), well above chance, and 89.9 (out of 139 admitted predictions, which is 58\% of all such claims) for claims modified to flip the label, but with unchanged evidence sentences. 

In practice with \textsc{BERT\textsubscript{base}+MemMatch}, one would deterministically know when the datastore has changed, so we could in principle make use of such information at inference, treating such instances with caution, as with special submodules or by requiring particularly confident predictions under the model.\footnote{In contrast, at inference, we cannot, in general, \textit{deterministically} know when the \textit{claim} has been altered to flip the label, or is otherwise an out-of-distribution instance, and instead must rely on the level distances or exemplar auditing.} For illustrative purposes, we include the result when we simply exchange the prediction with the opposite label on any instances falling outside the training mean and with modified evidence sentences (\textsc{BERT\textsubscript{base}+MemMatch+dist+exchange}). Interestingly, this raises the overall accuracy across all 717 claims to nearly 70. This is not a viable prediction strategy, in general, and in a real application one would produce null predictions or update the exemplar database, as discussed next, but it serves to illustrate that a non-trivial proportion of the unreliable predictions from that subset of the dataset are identifiable by large level 2 and 3 distances. Table~\ref{tab:retrieval-dist-examples-sym} in Appendix~\ref{sec:AppendixSymmetricAnalysis} provides examples of instances with close and far distances. 

In addition to the level distances, we can also use exemplar auditing to only admit a prediction if the nearest exemplar is a true positive, \textsc{BERT\textsubscript{base}+MemMatch+ExA\textsubscript{tp}}. We observe about a 1 point gain in accuracy across both \textsc{dev\textsubscript{1-evidence}} and \textsc{sym\textsubscript{gen.}} relative to the base inference approach, while generating predictions for about 90\% of the data. For reference, Figure~\ref{fig:exa-tp-dist} illustrates the fidelity of the distances to the true positive exemplars, with closer distances to the true positive exemplar associated with higher accuracy predictions.

The aforementioned results provide comparisons to previous work. We can use the newly released version of the data, which contains a development split, to demonstrate updating the model behavior by modifying the exemplar database. To put these results in perspective, in Table~\ref{table:symmetric-results-testv2} we provide the same inference approaches as those on the original split, and we see that the direction of the results is similar to those on \textsc{sym\textsubscript{gen.}} in Table~\ref{table:symmetric-results}. The overall higher accuracies are attributable to this split also including some of the original, unmodified claims from the \textsc{dev\textsubscript{1-evidence}} set.

With this new split, we can examine updating the exemplar database in the following way: We add both \textsc{train\textsubscript{1-evidence}} and \textsc{sym\textsubscript{dev.v2}} to the exemplar database, and at inference over \textsc{sym\textsubscript{test.v2}}, if we detect that the retrieval datastore has changed (i.e., the evidence sentence has changed relative to the original in Wikipedia), we ignore the model prediction and instead use the reference label associated with the nearest exemplar from \textsc{sym\textsubscript{dev.v2}}. Table~\ref{table:symmetric-results-testv2} includes this result as \textsc{BERT\textsubscript{base}+MemMatch+ExA\textsubscript{update}}. In this way, we make predictions over all instances, and the accuracy is also an improvement over the naive approach of exchanging the labels based on distances (\textsc{BERT\textsubscript{base}+MemMatch+dist+exchange}). The accuracy on the subset of claims associated with altered datastores rises from 52.5 to 67.1, demonstrating that we have substantially altered the behavior of the model, without updating the parameters via SGD updates in the traditional sense.

Qualitatively, exemplars with close distances exhibit shared abstract patterns with the claim and evidence of the matched test instances, as shown in Table~\ref{tab:exa-examples-tp} in Appendix~\ref{sec:AppendixSymmetricAnalysis}. Typically this occurs without overlap of the main entities (nouns), since FEVER is constructed to minimize the training and dev overlap of claims associated with the same Wikipedia article. By using the difference vectors from levels 2 and 3 as the exemplars, we can match against these more abstract patterns.

Each level of the model induces an alignment between the \textsc{query} sequence and the \textsc{support} sequence, resulting from the per-filter matching of the max-pooled CNNs across the two sequences. Figure~\ref{fig:ScaledLevel1Alignments} in Appendix~\ref{sec:AppendixParameterEstimates} illustrates this alignment. In some settings, this can serve as an additional means of analyzing the model and data; however, token-level alignments are in general difficult to reconcile with prediction labels with retrieval-classification tasks given the deep composition across tokens and sequences required. The level distances and exemplar auditing approach presented here provide an alternative means of analyzing such models.

\section{Discussion}\label{sec:discussion}

The relative success of our model points to broader applications. \textsc{BERT} has been shown to be effective on a wide variety of classification tasks. Our model can be frozen and augmented with an additional classification layer, such as \textsc{BLADE} \citep{Schmaltz-2019-BLADE-1} or \textsc{multiBLADE} \citep{Schmaltz2020-multiBLADE}, for use in similar binary or multi-label classification settings, but with the powerful addition of explicit retrieval.

FEVER is largely limited to one- or two-hop reasoning. For deeper multi-hop reasoning tasks, additional levels could be added commensurate with the nature of the search required for the given task. For a particularly deep search graph, both \textsc{query} and \textsc{support} sequences could be built-up as progressively constructed cross-sequences. In particular, we hypothesize that \textit{generative} language modeling can be recast in such a manner, inserting unsupervised or semi-supervised instances into the retrieval levels, which we leave for future work. 

\section{Additional Related Work}

There are a large number of additional multi-model, pipeline FEVER systems; we compared to a representative selection of full systems that have progressively pushed the state-of-the-art since the 2018 Shared Task. Pushing the upper-bound is beyond the scope of this paper, but we would assume that a larger Transformer, such as \textsc{BERT\textsubscript{large}} or GPT-3 \citep{BrownEtAl-2020-GPT-3}, would increase effectiveness, ceteris paribus.

Recent work suggests that dense (bi-encoder) representations can be effective in retrieval and sentence similarity tasks relative to traditional IR and domain-specific, feature-oriented models, at least when suitable pre-training is possible \citep{ChangEtAl-2020-Pretraining-Dense-IR}. \citet{HumeauEtAl-2020-Polyencoders} proposes an alternative architecture, poly-encoders, that is more effective than bi-encoders and more efficient, but less effective, than cross-encoders for sentence matching tasks. Our work is orthogonal to these works since our model addresses tasks that involve joint retrieval \textit{and} classification (i.e., there is a multi-stage search aspect requiring accumulation of disparate sequences). 

In NLP and ML, writ large, relatively few truly end-to-end\footnote{We take this \textit{informal} term to mean the neural model backpropagates jointly through both retrieval and classification, and inference is performed with that same single model.} neural models have been proposed for retrieval-classification tasks. The early work of \citet{YinAndRoth-2018-TwoWingOS} considers joint optimization of a binary vector over \textsc{support} sequences and a classification decision, with respect to a \textsc{query} sequence, via shared dense representations. Our work is conceptually distinct, recasting the problem in terms of non-parametric matching of memories from bi-encoders and cross-encoders. From the subfield of QA, the Open-Retrieval Question Answering system (ORQA) system of \citet{LeeEtAl-2019-ORQA-latent-retrieval} aims for end-to-end retrieval and classification; however, the means of doing so, and the learning setting, is fundamentally different than proposed and examined here. Whereas we restructure the model to directly accommodate search using a single shared Transformer LM with memory layers for the FEVER task, which has supervised evidence, ORQA relies on a special pre-training task for a bi-encoder retriever without supervised evidence, with a separate BERT model serving for classification. REALM \citep{GuuEtAl-2020-REALM} extends this work with a means of jointly updating the bi-encoder retriever with a separate Transformer classifier and additional pre-training strategies for QA tasks. In principle, unsupervised evidence can be accommodated in our training procedure by adding such instances to levels 1 and 2 (as for example, associating a sentence from the body of a Wikipedia article with one or more from the introduction, or similar sentence-level Cloze task), which we leave for future work.

\section{Conclusion}

We have presented a new, unified model that handles both retrieval and classification, using a single Transformer combined with memory layers, a supervised similarity loss, and a coarse-to-fine search procedure to access the non-parametric memory. With a modest additional amount of parameters relative to a standard pre-trained \textsc{BERT} model, we enable the language model to explicitly retrieve information, which we show significantly improves fact verification accuracy on \textsc{FEVER} compared to only relying on the language model parameters as a knowledge base. In this way, we significantly simplify the standard multi-model pipeline for retrieval-classification tasks, and we illustrate how tightly coupled retrieval and classification can be leveraged to identify low confidence instances, resulting from hypothesis bias, for example. We further extend exemplar auditing (non-parametric memory matching) to this sequence matching setting, illustrating a means of further analyzing the model and updating the model behavior.

\bibliography{memmatch}
\bibliographystyle{acl_natbib}

\clearpage
\appendix

\section{Alignment Visualization}

\begin{figure*}[hbt!]
    \centering
    \includegraphics[scale=0.50]{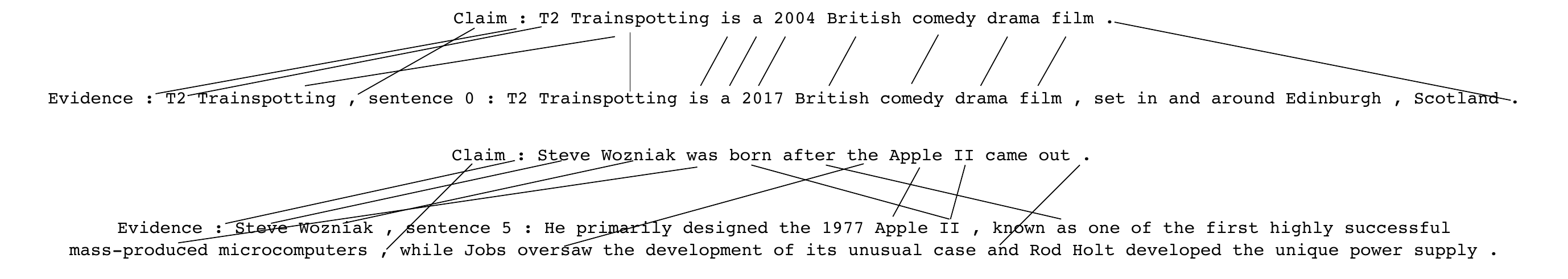}
    \caption{Max scaled alignment between the \textsc{query} and \textsc{support} sequences for level 1 (bi-sequence retrieval) for two claims, and their given evidence, from the \textsc{sym\textsubscript{test.v2}} dataset.}
    \label{fig:ScaledLevel1Alignments}
\end{figure*}

Each ``level'' of the model induces an alignment between the \textsc{query} sequence and the \textsc{support} sequence, resulting from the per-filter matching of the max-pooled CNNs across the two sequences. Figure~\ref{fig:ScaledLevel1Alignments} illustrates this alignment for two claims from the \textsc{sym\textsubscript{test.v2}} test dataset for level 1 (bi-sequence retrieval). For display purposes, we have scaled each per-token filter weight (of the \textsc{query} sequence) by the natural exponentiation of the negation of the filter difference, and summed these values for each \textit{(\textsc{query} sequence token index, \textsc{support} sequence token index)} pair, which uniquely describe an alignment in this convention. This simple scaling emphasizes tokens that have large activations that also have small filter differences. To further simplify the display here, we only show the maximum scaled pair for each \textsc{query} sequence token and then de-WordPiece-tokenize both sequences, merging shared alignments.

At the bi-sequence retrieval level, this often (but certainly not always) results in keywords being aligned across the \textsc{query} and \textsc{support} sequences. In some cases, this may be useful to analyze the model behavior in conjunction with the mechanisms described in the main text. However, some caution is in order in understanding these alignments, as token-to-token alignment is an underspecified task for fact verification (and QA-style classification)\footnote{Additionally, in the cross-sequence levels (levels 2 and 3), there is confounding in the input sequences to be aligned. If such unsupervised alignments are needed in such a setting, one possibility would be to mask the output tokens in the \textsc{support} sequence corresponding to the \textsc{query} sequence, but since such alignment is, in any case, underspecified in FEVER, we do not pursue that here.}, in general, given the deep cross token composition required. Furthermore, not apparent in the display simplification of Figure~\ref{fig:ScaledLevel1Alignments} is that the first four WordPieces\footnote{\texttt{C \#\#lai \#\#m :}} tend to be diffusely aligned to a long-tail of tokens. That is to say, these alignments for such retrieval-classification tasks are not directly actionable in the same sense as token-level contributions in sequence-labeling tasks \citep{Schmaltz-2019-BLADE-1}, and we must instead primarily rely on the mechanisms described in the main text for analyzing the model and data. 

\section{Parameter Estimates}\label{sec:AppendixParameterEstimates}

\textsc{BERT\textsubscript{base}+MemMatch} contains the parameters of \textsc{BERT\textsubscript{base}} (110 million) and those of the CNN and word embeddings of each level. With a vocabulary of 7500 and an embedding size of 300 for the randomly initialized non-contextualized WordPiece embeddings, and 1000 kernel-width-1 (``unigram'') CNN filters, this results in $((300+768)\cdot1000+300\cdot7500)$ parameters for each of the 3 levels (the hidden size of \textsc{BERT\textsubscript{base}} is 768). Not including the CNN bias weights, this results in 9,954,000 parameters, or roughly 120 million when also including \textsc{BERT\textsubscript{base}}.

The parameter count for \textsc{RAG} is from the original publication, with 220 million parameters from the two \textsc{BERT\textsubscript{base}} models of the bi-encoder retriever and 406 million parameters from the seq2seq \textsc{BART\textsubscript{large}} model used for generation (classification in the case of FEVER). The estimates for \textsc{CompoundLabel} and \textsc{NSMN} are from \citet{NieEtAl-2020-NSMN-style-label-sharing}. The \textsc{BERT\textsubscript{large}+KBfeat} model includes \textsc{BERT\textsubscript{large}}, a separate Named Entity Recognition model, and a separate textual entailment model, so it has at least the 340 million parameters of \textsc{BERT\textsubscript{large}}. 

Estimating the parameter counts of \textsc{GEAR} and \textsc{DREAM} is more complicated, but rough ranges are sufficient for the purposes here. In our parameter estimates, we cumulatively count all parameters for all the models used as part of the pipelines, including those of external linguistic tools. \textsc{GEAR} uses \textsc{BERT\textsubscript{base}} (110 million) as the encoder. As part of retrieval, it uses a constituency parser with ELMo embeddings, of which we estimate 93.6 million parameters are from ELMo alone, but we do not estimate the parser size. We also do not attempt to estimate the size of the model that is used in conjunction with the MediaWiki API for retrieval, nor the size of the actual graph network, and instead state the overall system has at least 204 million parameters. The \textsc{DREAM} pipeline uses \textsc{BERT\textsubscript{base}} (110 million) for semantic role labeling; \textsc{NSMN} (28 million) as part of retrieval; a RoBERTa cross-encoder for evidence selection, which is 125 million or 355 million parameters depending on whether the \textsc{base} or \textsc{large} version was used; and an XLNet cross-encoder for claim verification, which is 110 million or 340 million parameters depending on whether the \textsc{base} or \textsc{large} version was used. We do not attempt to estimate the size of the graph network, and instead estimate the overall count of the pipeline system as either contained in, or exceeding, the interval $\left(373~\text{million}, 833~\text{million}\right]$.


\section{Examples from the Symmetric Datasets}\label{sec:AppendixSymmetricAnalysis}

The analysis datasets of \citet{Schuster-Etal-2019-FEVER-symmetric-bias} are useful for examining out-of-domain detection and effectiveness. As noted in the main text, certain characteristics of the symmetric subsets make it relatively challenging for the model. Changes to the Wikipedia sentences are not restricted to alterations that might typically occur over time, but instead, may also change a fundamental characteristic of the entity. Table~\ref{tab:retrieval-dist-examples-sym} provides some illustrative examples from the \textsc{sym\textsubscript{gen.}} dataset in which the evidence sentences from Wikipedia have been modified to change the true facts. For example, in the second to last example, we see that the nationality of the American Caroline Kennedy is switched to Armenian. These type changes may be very challenging if the model has encoded such information into the parameters of the model. In a small number of cases, the updated datastore contradicts the Wikipedia title (which we re-introduce in preprocessing to match our format from training) due to parenthetical disambiguation information in the article title. Ideally, the model would identify such out-of-domain examples with large level 2 and/or 3 distances, as in the second and third examples. Finally, a more subtle challenge is that modifying Wikipedia sentences in isolation may introduce annotation artifacts that push the sentence to the tails of what might be typically seen in Wikipedia. For example, in the first example in the table, the phrase ``not known by any other name'' replaces the original ``known as El Libertador''. Such a construction would seem to be relatively unusual (at least anecdotally) for Wikipedia as the first sentence of a biographical entry (namely, if an individual is not known by another name, it is simply not explicitly mentioned). With this claim and evidence, the model exhibits a very large level 2 distance. 

Table~\ref{tab:exa-examples-tp} shows three correct predictions from \textsc{sym\textsubscript{test.v2}} with relatively low distance exemplars from \textsc{train\textsubscript{1-evidence}}. These are cases in which the Wikipedia sentences have not been modified. When the true positive exemplar distances are low, the mappings between the query and exemplar instances tend to exhibit similar abstract, relational patterns, despite often not having explicit overlap of the main entities. For example, in the first example, we see a similar mapping between the claim and evidence of a media show's setting (joined by ``is set in''); in the second example, a connection between two distinct individuals and their birth locations (``was born in''); and in the third example, both claims contain a statement about a politician with a mistake in the date.

\begin{table*}[t] 
\centering
\footnotesize
\begin{tabular}{P{35mm}T{105mm}}
\toprule
\rowcolor{lightestgray} & Claim id 1973810000004 (reference label: \textsc{false})\\
\multirow{3}{*}{\shortstack[l]{\textsc{support\textsubscript{level3}} sequence\\Level 2 distance: \textcolor{red}{162.77}\\Level 3 distance: 0.84}} & \ttfamily{\textcolor{red}{Supports:} Claim: Simón Bolívar is known as El Libertador. Evidence: Simón Bolívar, sentence 0: Simón José Antonio de la Santísima Trinidad Bolívar y Palacios (\string[\textipa{si"mon bo"liBar}\string]; 24 July 1783 -- 17 December 1830), \underline{not known by any other name}, was a Venezuelan military and political leader who played a leading role in the establishment of Venezuela, Bolivia, Colombia, Ecuador, Peru and Panama as sovereign states, independent of Spanish rule.}\\  
\midrule
\rowcolor{lightestgray} & Claim id 1314220000004 (reference label: \textsc{false})\\
\multirow{3}{*}{\shortstack[l]{\textsc{support\textsubscript{level3}} sequence\\Level 2 distance: \textcolor{red}{190.13}\\Level 3 distance: \textcolor{red}{1.44}}} & \ttfamily{\textcolor{red}{Supports:} Claim: Bones is a television series. Evidence: Bones (TV series), sentence 0: Bones is an American \underline{cuisine that has a lot of meat and dairy} \underline{products}.}\\ 
\midrule
\rowcolor{lightestgray} & Claim id 2024540000004 (reference label: \textsc{false})\\
\multirow{3}{*}{\shortstack[l]{\textsc{support\textsubscript{level3}} sequence\\Level 2 distance: \textcolor{red}{200.89}\\Level 3 distance: \textcolor{red}{2.39}}} & \ttfamily{\textcolor{red}{Supports:} Claim: Tinker Tailor Soldier Spy is an espionage film. Evidence: Tinker Tailor Soldier Spy (film), sentence 0: Tinker Tailor Soldier Spy is a 2011 \underline{music video} by Tomas Alfredson.}\\ 
\midrule
\rowcolor{lightestgray} & Claim id 1390370000004 (reference label: \textsc{true})\\
\multirow{3}{*}{\shortstack[l]{\textsc{support\textsubscript{level3}} sequence\\Level 2 distance: 0.05\\Level 3 distance: 0.54}} & \ttfamily{Supports: Claim: Star Trek: Discovery is an album. Evidence: Star Trek: Discovery, sentence 0: Star Trek: Discovery is an upcoming \underline{music album of Lady} \underline{Gaga}.}\\ 
\midrule
\rowcolor{lightestgray} & Claim id 630380000004 (reference label: \textsc{true})\\
\multirow{3}{*}{\shortstack[l]{\textsc{support\textsubscript{level3}} sequence\\Level 2 distance: 0.08\\Level 3 distance: 0.67}} & \ttfamily{Supports: Claim: Caroline Kennedy is Armenian. Evidence: Caroline Kennedy, sentence 0: Caroline Bouvier Kennedy (born November 27, 1957) is an \underline{Armenian} author, attorney, and diplomat who served the Ambassador to Japan from 2013 to 2017.}\\ 
\midrule
\rowcolor{lightestgray} & Claim id 72080000002 (reference label: \textsc{false})\\
\multirow{3}{*}{\shortstack[l]{\textsc{support\textsubscript{level3}} sequence\\Level 2 distance: 0.06\\Level 3 distance: \textcolor{red}{1.33}}} & \ttfamily{Refutes: Claim: Colombiana is a French film. Evidence: Colombiana, sentence 0: Colombiana is a 2011 \underline{Pakistani} action film co-written and produced by Luc Besson and directed by Olivier Megaton.}\\ 
\bottomrule
\end{tabular}
\caption{\label{tab:retrieval-dist-examples-sym} Examples (\textsc{support\textsubscript{level3}} sequences) of close (bottom 3) and far (top 3) level 2 (retrieval) distances from the \textsc{sym\textsubscript{gen.}} dataset in which the evidence sentences have been modified to \textit{change the true facts} (i.e., these are \textit{not} from the original FEVER dataset, and the new Wikipedia sentences have factually incorrect information). This dataset is useful for examining the model's out-of-distribution effectiveness and ability to identify outliers. We also show the level 3 (classification) distances. For reference, the mean level 2 distance from the training set at the top of the beam, given a correct retrieval, is 0.49 ($\pm 4.75$), and the mean level 3 distance, given a correct classification, is 0.92 ($\pm 1.80$). The bottom three instances are correctly predicted by the model, and the remainder have incorrect model predictions. Incorrect classification labels and distances beyond the training set mean are in red. We have \underline{manually underlined changes} made to the Wikipedia evidence sentences by the human annotators. The second and third instances illustrate the small number of cases in which our re-introduction of the title clashes with the annotator changes.
  }
\end{table*}

\begin{table*}[t] 
\centering
\footnotesize
\begin{tabular}{P{35mm}T{105mm}}
\toprule
\rowcolor{lightestgray} & Claim id 147493 (reference label: \textsc{true})\\
\multirow{2}{*}{\shortstack[l]{Test\\ \textsc{support\textsubscript{level3}} sequence}} & \ttfamily{Supports: Claim: T2 Trainspotting is set in and around a city. Evidence: T2 Trainspotting, sentence 0: T2 Trainspotting is a 2017 British comedy drama film, set in and around Edinburgh, Scotland.}\\
\multirow{3}{*}{\shortstack[l]{\textcolor{blue}{Exemplar} \\ \textsc{support\textsubscript{level3}} sequence\\Exemplar distance: 0.14}} & \textcolor{blue}{\ttfamily{Supports: Claim: All My Children is set in a fictional suburb of a city. Evidence: All My Children, sentence 1: Created by Agnes Nixon, All My Children is set in Pine Valley, Pennsylvania, a fictional suburb of Philadelphia, which is modeled on the actual Philadelphia suburb of Rosemont.}}\\
\midrule
\rowcolor{lightestgray} & Claim id 166634 (reference label: \textsc{false})\\
\multirow{2}{*}{\shortstack[l]{Test\\ \textsc{support\textsubscript{level3}} sequence}} & \ttfamily{Refutes: Claim: Anne Rice was born in Japan. Evidence: Anne Rice, sentence 5: Born in New Orleans, Rice spent much of her early life there before moving to Texas, and later to San Francisco.}\\
\multirow{3}{*}{\shortstack[l]{\textcolor{blue}{Exemplar} \\ \textsc{support\textsubscript{level3}} sequence\\Exemplar distance: 0.21}} & \textcolor{blue}{\ttfamily{Refutes: Claim: Emma Stone was born in Taiwan. Evidence: Emma Stone, sentence 5: Born and raised in Scottsdale, Arizona, Stone began acting as a child, in a theater production of The Wind in the Willows in 2000.}}\\
\midrule
\rowcolor{lightestgray} & Claim id 125398 (reference label: \textsc{false})\\
\multirow{2}{*}{\shortstack[l]{Test\\ \textsc{support\textsubscript{level3}} sequence}} & \ttfamily{Refutes: Claim: Harold Macmillan died on December 29, 1886. Evidence: Harold Macmillan, sentence 0: Maurice Harold Macmillan, 1st Earl of Stockton, (10 February 1894 -- 29 December 1986) was a British Conservative politician and statesman who served as the Prime Minister of the United Kingdom from 10 January 1957 to 19 October 1963.}\\
\multirow{3}{*}{\shortstack[l]{\textcolor{blue}{Exemplar} \\ \textsc{support\textsubscript{level3}} sequence\\Exemplar distance: 0.24}} & \textcolor{blue}{\ttfamily{Refutes: Claim: William McKinley served as a US president from March 4, 1887 until his death. Evidence: William McKinley, sentence 0: William McKinley Jr. (January 29, 1843 -- September 14, 1901) was an American politician and lawyer who served as the 25th President of the United States from March 4, 1897 until his assassination in September 1901, six months into his second term.}}\\
\bottomrule
\end{tabular}
\caption{\label{tab:exa-examples-tp} Three correct predictions from \textsc{sym\textsubscript{test.v2}} with relatively low distance exemplars from \textsc{train\textsubscript{1-evidence}}. In the last example, both sequences exhibit a date mismatch between the claim and the evidence. For display simplicity, we only show the \textsc{support} sequences from level 3. The exemplar vectors consists of the difference vectors between the \textsc{query} and \textsc{support} sequences from levels 2 and 3. 
  }
\end{table*}

\section{Coarse-to-Fine Search Example}\label{sec:AppendixSearchExample}

Table~\ref{tab:search-example} provides an example of the output from each level of the coarse-to-fine search for one claim from the Dev set. This also illustrates the input format of the bi- and cross- encoded sequences, including the format of the additional positive and negative instances used in training. 

\paragraph{Additional FEVER Details}

General fact verification is challenging, and to a certain extent, under-defined. To make the task sufficiently narrow for research purposes, FEVER imposes a number of synthetic constraints on data collection. In particular, claims are relatively short, declarative sentences; the decision label for the vast majority of claims can be determined by no more than two Wikipedia sentences; the Wikipedia ``articles'' only include the introductory sections; and there is at least some lexical overlap with the associated title of a relevant Wikipedia sentence and the claim in most instances. Nonetheless, it is a useful starting point. Qualitatively, the claims roughly have the flavor of arbitrary, well-formed output from a strong generative language model, ranging from correct statements, to statement with subtle factual mistakes (in dates, places, occupations, etc.), to obviously incorrect and nonsensical statements that are otherwise grammatically correct.

\begin{table*}[t] 
\centering
\tiny
\begin{tabular}{P{30mm}T{110mm}}
\toprule
\rowcolor{lightestgray} \multirow{1}{*}{\shortstack[l]{\textbf{Level 1}}} & \\ 
\multirow{1}{*}{\shortstack[l]{\textcolor{QueryColor}{\textsc{query}} sequence}} & \ttfamily{\textcolor{QueryColor}{Claim: Charles de Gaulle was a leader in the French Resistance.}}\\
\greysmallrule
\multirow{2}{*}{\shortstack[l]{\\\textsc{support} sequence,\\beam index 0}} & \ttfamily{Evidence: French Resistance, sentence 0: The French Resistance (La Résistance) was the collection of French resistance movements that fought against the Nazi German occupation of France and against the collaborationist Vichy régime during the Second World War.}\\
\greysmallrule
\multirow{2}{*}{\shortstack[l]{\textsc{support} sequence,\\beam index 1}} & \ttfamily{Evidence: Charles de Gaulle, sentence 1: He was the leader of Free France (1940 -- 44) and the head of the Provisional Government of the French Republic (1944 -- 46).}\\
\greysmallrule
$\vdots$ & \multicolumn{1}{c}{$\vdots$}\\
\greysmallrule
\multirow{2}{*}{\shortstack[l]{\textsc{support} sequence,\\beam index 14}} & \ttfamily{\textcolor{blue}{Evidence: Charles de Gaulle, sentence 12: Despite frosty relations with Britain and especially the United States, he emerged as the undisputed leader of the French resistance.}}\\
\greysmallrule
$\vdots$ & \multicolumn{1}{c}{$\vdots$}\\
\greysmallrule
\multirow{2}{*}{\shortstack[l]{\textsc{support} sequence,\\beam index 99}} & \ttfamily{Evidence: Resistance (EP), sentence 7: This EP or mini-album sold nearly all of its 200,000 copies.}\\
\midrule
\rowcolor{lightestgray} \multirow{1}{*}{\shortstack[l]{\textbf{Level 2}}} & \\ 
\multirow{1}{*}{\shortstack[l]{\textcolor{QueryColor}{\textsc{query}} sequence}} & \ttfamily{\textcolor{QueryColor}{Consider: Claim: Charles de Gaulle was a leader in the French Resistance.}}\\
\greysmallrule
\multirow{2}{*}{\shortstack[l]{\\\textsc{support} sequence,\\beam index 0}} & \ttfamily{Consider: Claim: Charles de Gaulle was a leader in the French Resistance. Evidence: Charles de Gaulle, sentence 1: He was the leader of Free France (1940 -- 44) and the head of the Provisional Government of the French Republic (1944 -- 46).}\\
\greysmallrule
\multirow{2}{*}{\shortstack[l]{\\\textsc{support} sequence,\\beam index 1}} & \ttfamily{Consider: Claim: Charles de Gaulle was a leader in the French Resistance. \textcolor{blue}{Evidence: Charles de Gaulle, sentence 12: Despite frosty relations with Britain and especially the United States, he emerged as the undisputed leader of the French resistance.}}\\
\greysmallrule
\multirow{2}{*}{\shortstack[l]{\\\textsc{support} sequence,\\beam index 2}} & \ttfamily{Consider: Claim: Charles de Gaulle was a leader in the French Resistance. Evidence: Charles de Gaulle, sentence 0: Charles André Joseph Marie de Gaulle (\string[\textipa{SAKl d@ gol}\string]; 22 November 1890 -- 9 November 1970) was a French general and statesman.}\\  
\midrule
\rowcolor{lightestgray} \multirow{1}{*}{\shortstack[l]{\textbf{Level 3}}} & \\ 
\multirow{1}{*}{\shortstack[l]{\textcolor{QueryColor}{\textsc{query}} sequence}} & \ttfamily{\textcolor{QueryColor}{Predict: Claim: Charles de Gaulle was a leader in the French Resistance.}}\\
\greysmallrule
\multirow{2}{*}{\shortstack[l]{\\\textsc{support} sequence,\\beam index 0}} & \ttfamily{Supports: Claim: Charles de Gaulle was a leader in the French Resistance. Evidence: Charles de Gaulle, sentence 1: He was the leader of Free France (1940 -- 44) and the head of the Provisional Government of the French Republic (1944 -- 46).  \textcolor{blue}{Evidence: Charles de Gaulle, sentence 12: Despite frosty relations with Britain and especially the United States, he emerged as the undisputed leader of the French resistance.} Evidence: Charles de Gaulle, sentence 0: Charles André Joseph Marie de Gaulle (\string[\textipa{SAKl d@ gol}\string]; 22 November 1890 -- 9 November 1970) was a French general and statesman.}\\
\greysmallrule
\multirow{2}{*}{\shortstack[l]{\\\textsc{support} sequence,\\beam index 1}} & \ttfamily{Unverifiable: Claim: Charles de Gaulle was a leader in the French Resistance. Evidence: Charles de Gaulle, sentence 1: He was the leader of Free France (1940 -- 44) and the head of the Provisional Government of the French Republic (1944 -- 46).  \textcolor{blue}{Evidence: Charles de Gaulle, sentence 12: Despite frosty relations with Britain and especially the United States, he emerged as the undisputed leader of the French resistance.} Evidence: Charles de Gaulle, sentence 0: Charles André Joseph Marie de Gaulle (\string[\textipa{SAKl d@ gol}\string]; 22 November 1890 -- 9 November 1970) was a French general and statesman.}\\
\greysmallrule
\multirow{2}{*}{\shortstack[l]{\\\textsc{support} sequence,\\beam index 2}} & \ttfamily{Refutes: Claim: Charles de Gaulle was a leader in the French Resistance. Evidence: Charles de Gaulle, sentence 1: He was the leader of Free France (1940 -- 44) and the head of the Provisional Government of the French Republic (1944 -- 46).  \textcolor{blue}{Evidence: Charles de Gaulle, sentence 12: Despite frosty relations with Britain and especially the United States, he emerged as the undisputed leader of the French resistance.} Evidence: Charles de Gaulle, sentence 0: Charles André Joseph Marie de Gaulle (\string[\textipa{SAKl d@ gol}\string]; 22 November 1890 -- 9 November 1970) was a French general and statesman.}\\
\midrule
\rowcolor{lightestgray} \multirow{1}{*}{\shortstack[l]{\textbf{Level 3 (training only)}}} & \\ 
\multirow{1}{*}{\shortstack[l]{\textcolor{QueryColor}{\textsc{query}} sequence}} & \ttfamily{\textcolor{QueryColor}{Reference: Claim: Charles de Gaulle was a leader in the French Resistance.}}\\
\greysmallrule
\multirow{2}{*}{\shortstack[l]{\\\textsc{support} sequence,\\positive training instance}} & \ttfamily{Supports: Claim: Charles de Gaulle was a leader in the French Resistance. \textcolor{blue}{Evidence: Charles de Gaulle, sentence 12: Despite frosty relations with Britain and especially the United States, he emerged as the undisputed leader of the French resistance.}}\\
\greysmallrule
\multirow{2}{*}{\shortstack[l]{\\\textsc{support} sequence,\\\textcolor{red}{negative} training instance}} & \ttfamily{\textcolor{red}{Refutes:} Claim: Charles de Gaulle was a leader in the French Resistance. \textcolor{blue}{Evidence: Charles de Gaulle, sentence 12: Despite frosty relations with Britain and especially the United States, he emerged as the undisputed leader of the French resistance.}}\\
\greysmallrule
\multirow{2}{*}{\shortstack[l]{\\\textsc{support} sequence,\\\textcolor{red}{negative} training instance}} & \ttfamily{\textcolor{red}{Unverifiable:} Claim: Charles de Gaulle was a leader in the French Resistance. \textcolor{blue}{Evidence: Charles de Gaulle, sentence 12: Despite frosty relations with Britain and especially the United States, he emerged as the undisputed leader of the French resistance.}}\\
\bottomrule
\end{tabular}
\caption{\label{tab:search-example} Examples of levels 1, 2, and 3 \textcolor{QueryColor}{\textsc{query}} and \textsc{support} sequences from a Dev set claim. If this were a training instance, we would also include the auxiliary \textsc{query} and \textsc{support} sequences in the final rows marked ``Level 3 (training only)'', creating positive and \textcolor{red}{negative} instances. (For training, in levels 1 and 2, the true instances would be the sequences with correct evidence, and the negative instances would be the nearest incorrect matches. In the prediction level 3, the true instance would be the cross-sequence with the correct label, and the two negative instances would contain the remaining incorrect labels.) The \textcolor{blue}{ground-truth evidence sentence (``Charles de Gaulle'', sentence 12)} rises from the 15th beam position in level 1 to the 2nd position in level 2, and it is included in the marginalization in level 3. Note the differing prefixes across sequences and levels, and that the \textsc{support} sequences in levels 2 and 3 are cross-encoded, while level 1 is bi-encoded. 
  }
\end{table*}

%
%


\section{Training Details}\label{sec:AppendixAdditionalTrainingDetails}

We train with separate optimizers for \textsc{BERT\textsubscript{base}}, for which we use Adam with weight decay \citep{LoshchilovAndHutter-2019-AdamW} as in previous works fine-tuning \textsc{BERT}, and the CNNs, for which we use Adadelta \citep{Zeiler-2012-Adadelta}. We alternately freeze either the Transformer parameters, or all of the CNN parameters, for each epoch, for all mini-batches in the epoch, which we found works well in practice. The intuition is that we seek for the CNN to produce distilled representations of the Transformer, conditional on the level, but if the gradients of both the Transformer and CNN parameters are updated at the same time, the high capacity lower layers of the Transformer could learn to fully describe the data, at the detriment of the quality of the representations of the memory layers, resulting in poor matching effectiveness during the non-parametric search procedure. 

The AdamW optimizer we use for \textsc{BERT} corresponds to BertAdam in the HuggingFace re-implementation of \citet{DevlinEtAl-2018-BERT}. We start the iterative freezing and unfreezing of the Transformer and CNNs by freezing the Transformer in the first epoch, and rotate thereafter every epoch. For training the CNNs (i.e., the memory layers), we use a learning rate of 1.0 with Adadelta. With BertAdam, we use a learning rate of $0.00002$ and a warmup proportion of 0.1. We freeze the Transformer after 6 epochs of updating the Transformer's parameters (i.e., 6 unfrozen epochs, or after epoch 12 overall); however, in practice, this only comes into effect in the initial training with $k_1=10$, during which the highest accuracy epoch is epoch 13, as when re-starting training with $k_1=30$, the highest accuracy epoch is epoch 2.

The non-parametric memories significantly exceed the memory capacity of the GPU, but they can be processed efficiently by caching to disk and performing the embarrassingly parallel distance calculations on the GPU. For the forward pass through the coarse-to-fine search, which does not require gradient accumulation, we can significantly reduce the overall search time, which is dominated by forward passes through the Transformer, by utilizing large batch sizes, performing each level of the search together across all claims. We use a batch size of 2.5k for level 1 and 1.2k for levels 2 and 3, with the resulting distance calculations processed in chunks of 4.8k to 50k, depending on the level. For backpropagating through the levels, we construct mini-batches of up to 9 claims, including all of the positive and hard negative sequences, across levels, for each of the claims in the mini-batch. 


\paragraph{Constructing Positive and Negative Training Instances} For verifiable claims, for level 1, we construct 1 or to 2 positive instances for each claim, corresponding to the 1 or 2 reference Wikipedia sentences, and 1 hard negative, which is the nearest incorrect match from the search procedure. For level 2, we construct 1 positive instance from the leading reference Wikipedia sentence (i.e., that which occurs earliest in the article) and 1 hard negative from the nearest incorrect match not in the reference evidence set. For level 3, we create a positive instance with the correct label and the top $z$ sentences from level 2, and we create 2 negative instances by flipping the classification label. For level 3, we also create 3 instances from the reference data (independent of search): A positive instance with the correct label and all correct reference Wikipedia sentences, and 2 negative instances by flipping the classification label. 

Unverifiable claims do not have sequences corresponding to levels 1 and 2 in training (since there is no ground-truth reference evidence), nor the reference level 3 sequences\footnote{However, as noted above, the unverifiable label does occur as a hard negative for verifiable claims in the reference instances of level 3.}, but the prediction level 3 sequences for unverifiable claims are constructed in a manner analogous to the verifiable claims. In principle, additional instances could be created---as for example, adding additional hard negatives---at a concomitant cost of additional training time. We find that the aforementioned constructions yield strong results, in relative terms, and keep training manageable on a single GPU. 

\end{document}